\documentclass[12pt,preprint]{aastex}
\usepackage{color}

\shorttitle{The largest oxigen bearing organic molecule repository}
\shortauthors{Requena-Torres et al.}

\begin{document}


\title{The Galactic center, the largest oxigen bearing organic molecule repository}

\author{Requena-Torres, M.A.\altaffilmark{1}\email{requena@damir.iem.csic.es}, Mart\'in-Pintado, J.\altaffilmark{1}\email{martin@damir.iem.csic.es},\\
Mart\'in, S.,\altaffilmark{2}\email{martin@astro.cfa.edu} \& Morris, M. R.\altaffilmark{3}\email{morris@astro.ucla.edu}}

\altaffiltext{1}{Departamento de Astrof\'isica Molecular e Infrarroja-
Instituto de Estructura de la Materia-CSIC, C\ Serrano 121, E-28006 Madrid, Spain}
\altaffiltext{2}{Harvard-Smithsonian Center for Astrophysics, 60 Garden Street, MS 78 Cambridge, MA 02138, USA}
\altaffiltext{3}{Department of Physics \& Astronomy, University of California, Los Angeles, CA 90095-1547, USA}

\begin{abstract}
We present the first detection of complex aldehydes and isomers in three
typical molecular clouds located within 200$\,$pc of the center of our Galaxy.
 We
find very large abundances of these complex organic molecules (COMs) in the
central molecular zone (CMZ), which we attribute to the ejection of COMs from grain mantles
by shocks. The relative abundances of the different COMs with respect to that
of CH$_3$OH are strikingly similar for the three sources, located in very different
environments in the CMZ. The similar relative abundances point toward a unique
grain mantle composition in the CMZ. Studying the Galactic center clouds and
objects in the Galactic disk having large abundances of COMs, we find that
more saturated molecules are more abundant than the non-saturated ones. We also
find differences between the relative abundance between COMs in the CMZ and
the Galactic disk, suggesting different chemical histories of the grain 
mantles between the two regions in the Galaxy for the complex aldehydes. 
Different possibilities for the
grain chemistry on the icy mantles in the GC clouds are briefly discussed. 
Cosmic rays can play an important role in the grain chemistry. With these new
detections, the molecular clouds in the Galactic center appear to be one of the
best laboratories for studying the formation of COMs in the Galaxy.
\end{abstract}

\keywords{astrochemistry --- molecular data --- ISM: molecules --- 
techniques:spectroscopic --- ISM:individual(CMZ)}
\section{Introduction}

It has been found that complex organic molecules (COMs) are very abundant in 
the Galactic center \citep[][hereafter Paper I]{marpin01,req06},
where the molecular gas is concentrated in the Central Molecular Zone
\citep[CMZ,][]{mor96}.
Large COMs in the Galaxy have been mainly associated with hot cores 
\citep[e.g.][]{ike01}, where they are evaporated from icy mantles of warm dust 
($>$100$\,$K) heated by massive proto-stars. The physical conditions in the CMZ 
are very different from those in hot cores. High gas-kinetic temperatures of 
$>$100~K, but low excitation temperatures $\sim$10--20~K
due to the relatively low H$_2$ densities (few 10$^{4}$ cm$^{-3}$) and cold 
dust of T$_{d}$$<$20$\,$K \citep[][]{nem00} are typical in the CMZ. 
Since the dust is too cold for
the evaporation of the ice-mantles, it has been proposed that the sputtering 
of the grains and grain mantles produced by widespread shocks is responsible 
for the rich chemistry and
the high temperatures observed in the CMZ \citep[][Paper I]{marpin97,hut98,che03}.
In spite of the different physical conditions in
hot cores and GC clouds, Paper I shows that the abundances of the COMs -C$_2$H$_5$OH,
CH$_3$OCH$_3$, HCOOCH$_3$, and HCOOH- relative to 
that of CH$_3$OH (methanol) are surprisingly similar for both
types of sources, suggesting a sort of ``universal'' grain mantle composition 
in the Galactic disk and in the CMZ .\\ 

Recently, even larger COMs like the aldehydes
propynal (HC$_2$CHO), propenal (CH$_2$CHCHO), propanal (CH$_3$CH$_2$CHO), and 
glycolaldehyde (CH$_2$OHCHO), and complex alcohols like ethylene glycol 
(HOCH$_2$CH$_2$OH) have been detected towards one star forming region in the
CMZ, Sgr~B2N \citep{dic01,hol00,hol02,hol04b}. The proposed formation 
paths for the COMs on grains include the hydrogenation, oxidation and/or carbon
addition of 
small molecules on grain mantles \citep{tie82,cha05}. 
Carbon addition on grain mantles has been studied by 
\citet{jam06} but it is unclear if these reactions are as efficient as
predicted. \citet{hir04, wat05} and Fuchs et al. (2007, in press) have shown 
that the 
hydrogenation of CO can efficiently produce H$_2$CO and CH$_3$OH. 
Reactions of radicals on 
grain surfaces constitute another hypothesis for forming large molecules 
\citep[][Garrod et al., in prep]{gre76,gar06}. 
Recently \citet{ben07} have formed the isomers 
CH$_2$OHCHO and HCOOCH$_3$ on grain mantle analogs by irradiation of a mixture 
of CH$_3$OH and CO ices at 10$\,$K with energetic electrons to mimic the effects 
of the cosmic rays. 
Finally, \citet{hal06} showed that in the Sgr~B2N hot core, 
the formation of CH$_2$OHCHO
from H$_2$CO in the gas phase may be possible. So far, most of our knowledge of 
these COMs is restricted to one source, the Sgr~B2N hot core, which has a very
complex core-halo structure \citep{marpin90,hol04a} making it very difficult 
to establish where (hot core versus envelope) these molecules are formed.\\

In this paper, we present the first detections of most of the largest COMs, 
previously observed only toward Sgr~B2N, in typical 
molecular clouds in the CMZ without signatures of star formation. 
These observations show that the Galactic center is one of the best 
laboratories for understanding the chemistry of COMs and support the idea of a 
similar grain mantle composition throughout the CMZ but this does not extend to
the Galactic disk.

\section{Observations and results}
Observations were carried in April and May of 2006 with the NRAO 
\footnote{The National Radio Astronomy Observatories is a facility of the National
Science Foundation, operated under a cooperative agreement by Associated
Universities, Inc.} 
100-m Robert C. Byrd Green Bank Telescope in West Virginia (USA). 
Ku-band and K-band receivers were used in connection with the spectrometer 
divided into four spectral windows of 200$\,$MHz each in
both bands. We obtained a resolution of 24.4$\,$kHz, 0.6$\,$km s$^{-1}$ for Ku 
band and 0.3$\,$km s$^{-1}$ for K band. Two polarizations were observed for
each spectral window through the use of offset oscillators in the IF.
The frequencies and the spectroscopic parameters of the observed molecular
transitions 
are shown in Table \ref{freq}. They were obtained from the Cologne Database for 
Molecular Spectroscopy \citep[][]{mul01,mul05}, the Jet Propulsion Laboratory 
catalog \citep[][]{pic98} and from \citet{hol04b}. 
    
The observed typical molecular clouds are MC~G--0.11--0.08 (the ``20 km s$^{-1}$
cloud''), and MC~G--0.02--0.07 (the ``50 km s$^{-1}$ cloud'') located in 
Sgr~A$^*$ complex, and MC~G+0.693--0.03 
 located in the  Sgr~B2 complex. Those sources  
were selected from Paper I due to their large column density of COMs.  In 
Table \ref{poss} we give for the observed sources the position, the velocity, 
the H$_2$ column density (derived from the
C$^{18}$O emission in Paper I), and the CH$_3$OH abundances also from Paper I.
The observations were made using the position-switched mode, with the off 
position free from emission selected from 
CS maps \citep{bal87}. 
The two polarization outputs from the spectrometer were
averaged in the final data reduction process to improve the signal-to-noise
ratio. After smoothing the data, the final resolution was $\sim$3$\,$km s$^{-1}$.
Calibration was achieved by using a noise tube, and the line intensities are 
given in the $T^{*}_{\rm A}$ scale (estimated with 10\%--20\% uncertainties), 
appropriate for extended sources.

\begin{figure}[h!]
\centering
\includegraphics[angle=0,width=11cm]{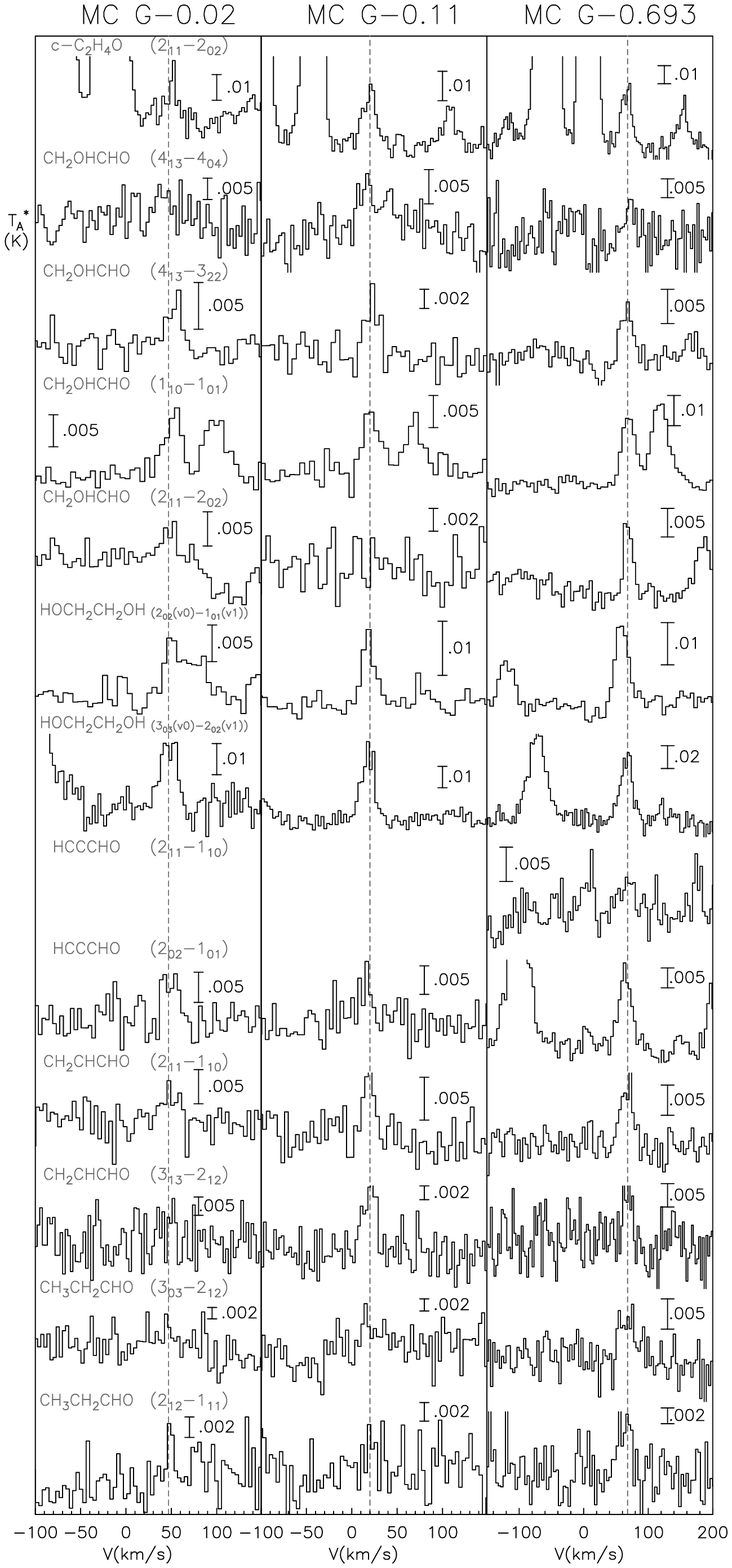}
\caption{Spectra of the molecular transitions  
observed in all the three sources. 
\label{spec}}
\end{figure}

\begin{figure}[h!]
\centering
\includegraphics[angle=270,width=13cm]{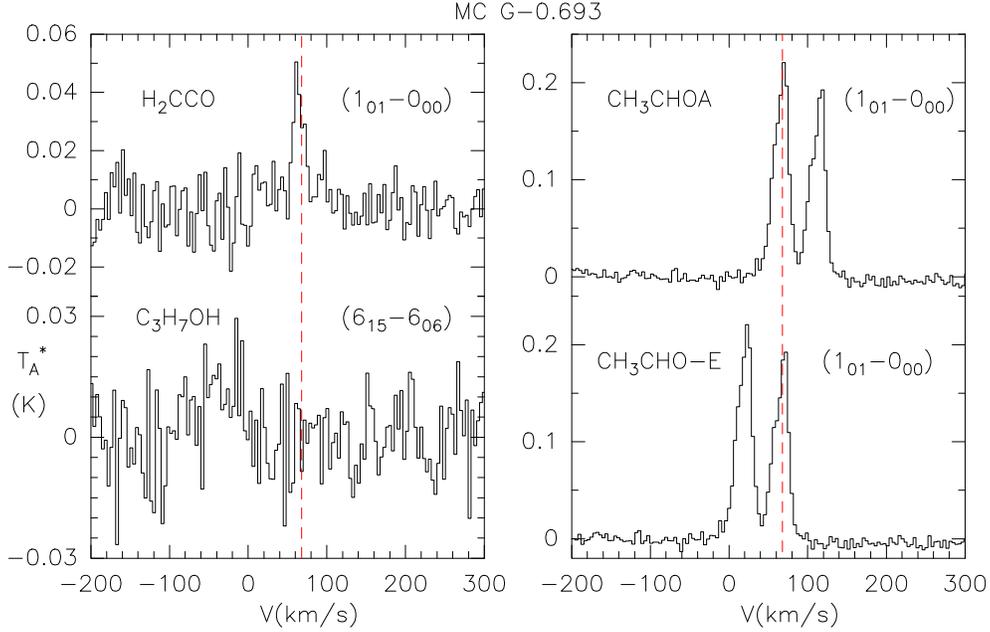}
\caption{Spectra of the extra molecular transitions 
observed in MC~G+0.693--0.03. This is the source with the most detections. 
\label{spec2}}
\end{figure}

Fig. \ref{spec} shows a sample of the line profiles observed toward the
different sources. We obtained detections of almost all the 
transitions in Table \ref{poss}. In MC~G+0.693--0.03 we were able to obtain the most 
complete set of 
transitions, including the HCCCHO 2$_{1,1}$$\rightarrow$1$_{1,0}$ line and the
extra molecular transitions shown in Fig. \ref{spec2}. In this source all the
transitions were detected for all molecules except for propanol (C$_3$H$_7$OH). 
As shown in Figs. \ref{spec} and \ref{spec2}, together with the complex 
aldehydes we also detect or derive upper limits for 
other related molecules like CH$_3$OH, HCOOCH$_3$ (methyl formate), CH$_3$CHO 
(acetaldehyde), H$_2$CCO (ketene), and c--C$_2$H$_4$O (ethylene oxide). 
We have also detected, H$_2$COH$^+$, protonated formaldehyde, 
one of the molecules expected to play a key role in the 
gas-phase formation of CH$_2$OHCHO in hot cores 
\citep{ohi96}. For completeness, we have also included 
data of H$_2^{13}$CO and H$_2$C$^{18}$O from \citet{mar06}, 
to derive the abundance of the chemically related molecule H$_2$CO. 

\section{Analysis}
\begin{figure}[h!]
\centering
\includegraphics[angle=270,width=15cm]{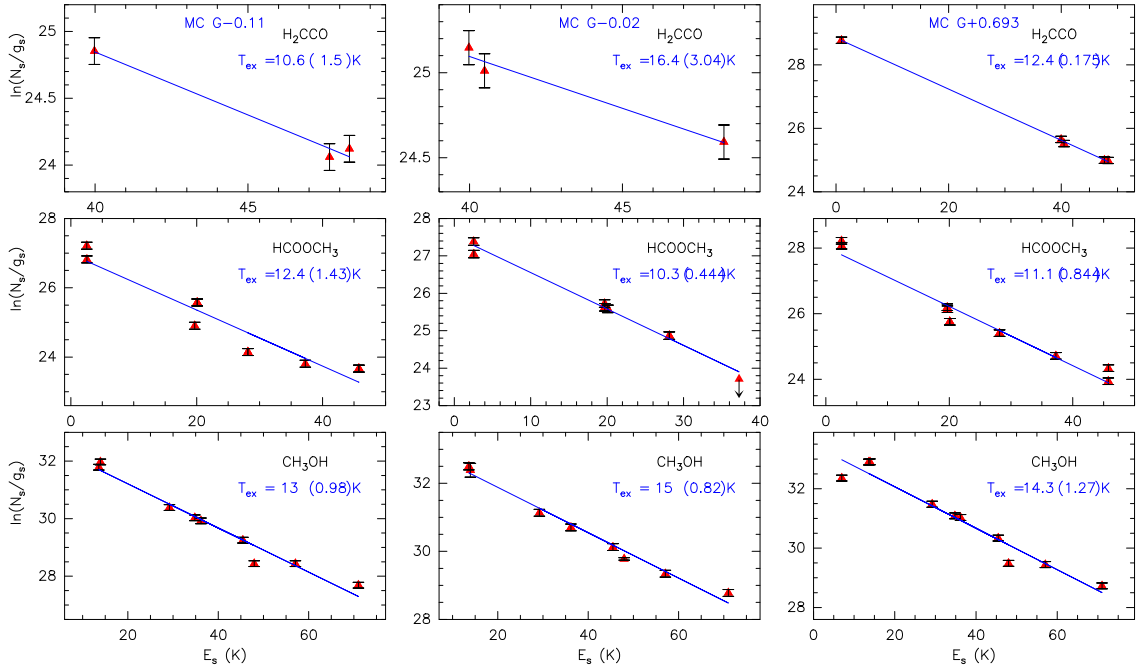}
\caption{Population diagrams for H$_2$CCO, HCOOCH$_3$ and CH$_3$OH in 
the three observed sources.
The population diagrams of CH$_3$OH and HCOOCH$_3$ also include data
from Paper I. 
\label{trot}}
\end{figure}

We obtained the line parameters shown in Table 
\ref{gauss} by fitting Gaussians to the observed line 
profiles. The column densities for all the molecules where then derived using 
the Local Thermodynamic Equilibrium (LTE) approximation, and the 
excitation temperatures, $T_{\rm ex}$, were derived from the population 
diagrams when enough rotational lines were available. 
Fig. \ref{trot} shows a sample of population diagrams for the three
sources. We derived $T_{\rm ex}$ between 8--16$\,$K for all the sources, similar
to those found in Paper I. \\

The population diagrams for CH$_2$OHCHO cannot be fit with a single 
excitation temperature, indicating non-LTE excitation. This is 
consistent with the findings of \citet{hol04a}, who observed the
2$_{1,1}$$\rightarrow$2$_{0,2}$ and the 4$_{1,3}$$\rightarrow$3$_{0,4}$ lines 
in absorption and the 1$_{1,0}$$\rightarrow$1$_{0,1}$ in 
emission toward the Sgr~B2N hot core. For our analysis we have used the 
1$_{1,0}$$\rightarrow$1$_{0,1}$ and 4$_{1,3}$$\rightarrow$3$_{2,2}$ emission lines,
which yield similar column densities and do not show absorption profiles toward
Sgr~B2N. \\

The main source of error in our estimated column densities arises from the 
assumption of the LTE approximation using the single excitation temperature 
derived from the 
population diagrams. However, even in the extreme case that the excitation 
temperature varies by a factor of 2, the error introduced by these uncertainties
in our estimated column densities for COMs 
will be less than a factor of 2. We therefore consider that our estimation of 
column densities is uncertain by less than a factor of 2 (see also Paper I).
To obtain 
the relative molecular abundances, we have used the H$_2$ column densities 
derived in Paper I from the C$^{18}$O emission (see Table \ref{poss}). 
The results of the relative abundances of COMs are shown in Table 
\ref{abun}.

\section{Excitation}

As discussed in Paper I, 
the COMs in the GC show very low $T_{\rm ex}$ as compared with those
found in hot cores. This is in contrast with the gas kinetic temperatures 
derived from NH$_3$ and H$_2$ which range from 30 up to 500$\,$K \citep{hut93,nem01}. 
On average, 30\% of the molecular gas in the GC has kinetic temperatures of
$\gtrsim$100$\,$K. From the analysis of all the data, \citet{nem01} concluded that
the large abundances of NH$_3$, SiO, and the high kinetic temperatures but 
low dust temperatures can be explained by C--type shocks with velocities of
10$\,$km s$^{-1}$. The large COM abundances can also be explained by grain 
mantle sputtering by shocks. Shocks with velocities of $\sim$6$\,$km s$^{-1}$ are 
very efficient at ejecting molecules from the icy mantles when CO is considered
as the sputtering agent \citep{jimenez07}. \\

The most likely
explanation for the low $T_{\rm ex}$ derived from the various COMs is that the 
gas has high kinetic temperatures, but the COMs are subtermally excited due to
the relatively low H$_2$ column densities. This has been discussed in detail in 
Paper I with a similar
analysis of the excitation of CH$_3$OH. Furthermore, if the gas kinetic
temperature were on the order of 10-20$\,$K, in equilibrium with the cold dust,
then the COMs would be thermalized, requiring H$_2$ densities of 
$\sim$10$^6$ cm$^{-3}$ (Paper I). From the H$_2$
column densities in Table \ref{poss} the derived size of the molecular cloud
along the line of sight would be only $\sim$4.1$\times$10$^{16}$cm, just
2739$\,$AU. This is unlikely since the COM emission is very extended so this
would require that the GC molecular clouds be very thin sheets perpendicular to
the line of sight.

\section{Abundances and relative abundances}
We have found very high abundances for all the observed COMs in the CMZ, from 
10$^{-10}$ to 10$^{-8}$ except for CH$_3$OH,
which shows larger relative abundances of a few times 10$^{-7}$--10$^{-6}$. 
In our previous work on COMs in the CMZ (Paper I) 
we suggested that this region is affected by the
erosion of the dust grain mantles by shocks, and that the
gas-phase abundances of large COMs may be large if they are formed
on grain mantles. \\

To illustrate the possible differences in chemistry, we present the 
abundances of the different molecules as ratios with respect that that of CH$_3$OH in 
Table \ref{abun2} (as in Paper I). Errors in
the abundance ratios can be estimated by considering the propagation of the 
errors in the 
column densities in each molecule. Uncertainties of a factor of $\lesssim$3 
are expected for these abundance ratios. However, as mentioned in the previous
section the main source of uncertainty is the derived $T_{\rm ex}$, which will 
affect all COMs and CH$_3$OH in a similar way. In this case a factor of 3 must be
considered an upper limit to the uncertainties in our estimated abundance ratios.
The relative abundance of every molecule with respect to that of CH$_3$OH is 
rather constant between sources, with changes by only a factor of $\lesssim$3, 
similar to the uncertainties. As in
Paper I, we find similar abundance ratios for all COMs between sources that are separated
by hundreds of parsecs. This remarkable result supports the claim 
of a similar grain mantle composition in the whole CMZ.

\section{Comparison between related molecules}

Fig. \ref{oxi} summarizes all the observed COMs in the CMZ sources 
in this paper and in 
Paper I following the simplest hypothesis that the formation of COMs on grain 
mantles is due to additions of O, C and H to CO \citep[][]{tie82,cha05}. 
Chemical complexity starting from CO increases from left to right by adding 
C, from the center to top and bottom by adding H and in 
diagonals by adding O. 
The radicals are shown within ellipses and the molecular species in boxes. 
The different families of isomers are shown in the figure enclosed in
dashed line boxes. The molecules detected in the GC molecular
clouds are highlighted in boldface to distinguished them from the undetected species.\\
To test the scheme proposed in Fig. \ref{oxi}, we have compared the 
relative abundances derived between related molecules (see Table \ref{fam}),
like the isomers and the aldehyde/alcohol pairs between the CMZ sources,
 the star forming region Sgr~B2N and a sample of hot cores in the
 Galactic disk 
 \citep{bis07}.\\ 

\subsection{Isomers}

We find that the ratios between the different isomers
in the CMZ sources are very similar within a factor of $\sim$3.
Only the (CH$_3$)$_2$O:C$_2$H$_5$OH ratio shows some dispersion between
sources in the CMZ.
For the isomers c--C$_2$H$_4$O:CH$_3$CHO ratio, the CMZ sources show 
ratios that are in agreement with those measured in hot cores 
\citep{num98,ben05}. However, for
the (CH$_3$)$_2$O:C$_2$H$_5$OH ratio, we find a large difference between
the GC clouds and the hot core sources by nearly a factor of 10. The 
ratios CH$_3$COOH:CH$_2$OHCHO:HCOOCH$_3$, where we have only upper limits for the
CH$_3$COOH abundance, is relatively constant for the CMZ sources, but the abundance of 
CH$_2$OHCHO with respect to that of HCOOCH$_3$ is larger in the CMZ than in hot
cores by a factor of 10. The trend observed in the relative abundances between
isomers indicates a 
similar grain mantle composition in the CMZ, but different than in hot core 
regions. The new detections of COMs 
clearly show that the chemistry in the CMZ differs substantially from
that in hot cores in the Galactic disk.\\

\subsection{Aldehyde/alcohol pair}
Previous observations of the aldehyde/alcohol pairs in hot cores have shown 
that the COM chemistry favors the reduced alcohols with respect to their 
corresponding aldehydes \citep{ike01,hol02,bis07}, i. e. the double bond
between the C and the O must be easily broken. In the CMZ we found a similar 
trend (Table \ref{fam}). However, when we compare the hot cores with the CMZ 
clouds we find very large differences between their relative abundances. For the
H$_2$CO:CH$_3$OH and the CH$_2$OHCHO:HOCH$_2$CH$_2$OH pairs we find small 
differences and it could be explained by slightly different grain mantle 
composition between CMZ sources and hot cores. However, for the 
CH$_3$CHO:C$_2$H$_5$OH pair, the ratio
in the CMZ is more than 2 orders of magnitude larger than in hot cores. 
This might be attributable to the suggestion that the CH$_3$CHO formation in hot cores
may be related with a cooler gas in the external part of the hot core 
\citep{bis07}.\\ 

\subsection{Degree of saturation}
Finally, we compare the abundances between molecules related by a different
degree of saturation, breaking double or triple bonds (columns in Fig.
\ref{oxi}).
For the pair H$_2$CCO:CH$_3$CHO the relative abundances observed in the CMZ 
are completely different than observed in hot cores. The relative CH$_3$CHO 
abundance with respect to that of H$_2$CCO is larger by more that one order of 
magnitude in the CMZ clouds than that observed by \citet{bis07} in hot cores.
The break of the double bond seems to
be produced more efficiently in the CMZ
, maybe 
because of a more efficient hydrogenation. 
For the aldehydes bearing 3 carbons, we found similar ratios 
between CMZ sources. 
Unfortunately, these aldehydes have only been detected toward
one hot core, Sgr~B2N. Moreover, we derive similar abundance ratios in the 
Sgr~B2N hot core than in the other CMZ clouds for these molecules, assuming a 
low excitation temperatures \citep[10$\,$K, ][]{hol04b}.
Their extended emission in the CMZ is a clear indication that these molecules 
might not arise from the hot core, but from the envelope \citep{marpin90}. 
Detection of these molecules in other hot cores is required to establish whether 
the addition of carbon also occurs with the same efficiency in hot cores as in 
the CMZ.\\

\section{The formation of COMs}
The formation paths for the observed COMs are
not completely understood, either in the gas phase or on grain mantles 
\citep[][]{cha05, hal06}. 
From the observational point of view, up to now, our knowledge of the most 
complex organic molecules has been restricted to just one position toward a hot 
core, Sgr~B2N, with several velocity components that complicate its study. 
The detection of very
similar abundances of large COMs in several different zones in the CMZ opens the
possibility of constraining the formation mechanism for the most complex
molecules found in the interstellar medium. \\

\subsection{Gas phase}
The very high abundances obtained for the COMs and again the similar 
abundance ratios between the different COMs with respect to that of 
CH$_3$OH strongly supports the hypothesis that the same chemistry
occurs at different locations in the CMZ. Gas phase reactions
of simple organic molecules ejected from grain mantles to form the more complex
organic molecules \citep[as proposed by ][]{cha92} are unlikely to
produce the same, uniform abundance ratios, because the abundances of the
daughter species are strongly time dependent and also depend on the temperature and
density.

The detection of H$_2$COH$^+$ towards the Sgr~B2N hot core has been
proposed as an indication of the gas-phase formation of CH$_2$OHCHO 
\citep{hal06}.
However, in the CMZ the abundance of H$_2$COH$^+$ seems to be very low as
compared with that of CH$_2$OHCHO (Table \ref{fam}) to explain the formation of
this molecule in gas phase. 
The most likely explanation is that CH$_2$OHCHO, like the other COMs, is formed 
on grain mantles.\\

We stress that the GC molecular clouds are likely to be more affected by
different energetic processes, UV radiation, X-rays, supernova remnants, 
shocks, cosmic
rays than the clouds in the Galactic disk. These effects can drive the chemistry in grain mantles
and/or increase the available H in the gas phase or in icy grain mantles, making 
possible a larger degree
of hydrogenation in the CMZ. Fast hydrogenation reactions in the 
gas-phase after the ejection of COMs from the mantles could be also possible, 
but should affect 
differently the different clouds. In particular, such a situation should be time
dependent since the degree of hydrogenation will depend on the time
scales when the parent COMs were ejected from the grain mantles. As discussed
in Paper I, we do not find any change in the relative abundances of fully saturated
molecules like the C$_2$H$_5$OH, suggesting that gas phase chemistry after the 
ejection of molecules from grain mantles does not seems to be a fundamental 
process.\\

\subsection{Icy mantles of grains}
The constant relative abundances and abundance ratios of the more complex 
organic molecules observed in the 
CMZ indicate that the chemistry of all these molecules is likely to be related with 
common processes taking place throughout the CMZ. The formation of those complex 
molecules by hydrogenation of CO to form H$_2$CO and   
CH$_3$OH on grain mantles seems to be very efficient
in laboratory experiments \citep[][ and Fuchs 2007, in press]{hir04,wat05}. Similar 
hydrogenation reactions might then be possible for the formation of more 
complex molecules by addition of O and C (see Fig. \ref{oxi}). In fact, one of the 
largest COMs, the HOCH$_2$CH$_2$OH has even been detected in
comets by \citet{cro04}.\\

In the scheme proposed in Fig. \ref{oxi}, chemical complexity follows from the
availability of atomic C, H and O in the gas phase. The C availability results from
cosmic ray ionization and primary and secondary electron excitation of 
H$_2$ followed by fluorescence in the Lyman-Werner bands that produce FUV photons
capable to photodissociate CO. The fractional C abundance is 
C/CO$\sim$6$\times$10$^{-3}$, largely independent of the cosmic ray ionization
rate and therefore the carbon addition has to be similar in the different regions
\citep{tie05}.
Other formation routes from various simple molecules, 
including hydrocarbons such as acetylene (C$_2$H$_2$), where its triple bond 
is readily attacked by atomic H, could also lead to COMs such as 
CH$_2$CHOH as well as C$_2$H$_5$OH. \\

Another attractive alternative for the formation of the largest COMs on grains 
is that energetic processes in the GC fragment the simple molecules formed
by hydrogenation of CO like, H$_2$CO and CH$_3$OH triggering further 
hydrogenation and other reactions on grain surfaces. 
Following this hypothesis for the formation of COMs, the reactions of 
radicals at warm dust temperatures \citep{gar06} seems to be important, but the 
present chemical models do not show any result for the complex aldehydes we 
have observed in this work and the dust temperatures in the CMZ are relatively
low, only 10--20$\,$K. Several COMs, such as the simpler
amino acids, can be formed in the solid phase from H$_2$O, CH$_3$OH, NH$_3$ and CO or 
HCN ices affected by photochemistry \citep{dwo01,ber02}. However, most of the
COMs are easily photodissociated by the UV radiation needed to drive the
chemistry (see Paper I).\\

A very interesting alternative has been investigated by  
\citet{ben07} who have irradiated a 
mixture of CH$_3$OH and CO ices at 10$\,$K with energetic electrons to mimic the 
effects of MeV cosmic rays. As previously mentioned, GC 
clouds could be affected by an enhanced flux of cosmic rays. Cosmic rays  
break a C--H bond of CH$_3$OH, generating the hydroxymethyl (CH$_2$OH) and the 
metoxy
(CH$_3$O) radicals, plus H. This H has sufficient energy to overcome barriers
on the grain mantles to hydrogenate CO to form HCO. This radical will recombine 
with CH$_2$OH to form CH$_2$OHCHO and with CH$_3$O to form HCOOCH$_3$. In this
scheme, the isomers, CH$_2$OHCHO  
and HCOOCH$_3$, are efficiently formed while their other isomer CH$_3$COOH was 
produced with less efficiency than the other two.
The same trend predicted by this experiment in the abundances of these isomers 
is observed in the GC clouds.
Furthermore, HCOOCH$_3$ is formed more efficiently that CH$_2$OHCHO by one order
of magnitude,
just as we observed in the GC clouds. This suggests than cosmic rays
can be the driver of grain chemistry in the GC. However, it is unclear if the
estimated abundances of the C$_2$H$_4$O$_2$ species on the ices of 1\% of water
abundances \citet{ben07} is 
large enough to account for the abundances estimated
from our observation. This would require that the CH$_3$OH abundance in the 
mantles be similar to that of water in the GC. \\

All these arguments emphasize the suggestion made in Paper I that we observe in the
CMZ a uniform grain mantle composition ejected directly from the grain mantles 
to the gas
phase. However, the grain chemistry in the CMZ differs substantially 
from that in hot cores.\\

\section{Potential of the CMZ for future searches of COMs}
Because of the large number of detections of COMs in the 
CMZ, as well as their large abundances, this region stands out as one of the promising  
sources for detecting new molecules and for studying the origin of COMs in 
the Galaxy. In hot cores, the other prolific sources of COMs, 
line confusion and large partition functions make the detection/identification 
of very large molecules very difficult. 
In the CMZ, the COMs are widespread and show low excitation temperatures 
($\lesssim$20$\,$K), partially avoiding the large partition function and the extreme 
line confusion problems of the hot cores. 
Moreover, there are molecules like the complex aldehydes which are more 
abundant in the CMZ clouds affected by shocks than in hot cores.
Observations of these molecules in a larger sample of CMZ clouds and hot cores, 
and new studies of the chemistry for the large aldehydes are necessary to
clearly establish the different history of the grain mantles in the GC and in
the disk.

\section{Conclusions}
We have detected the most complex organic molecules, 
HC$_2$CHO, CH$_2$CHCHO, CH$_3$CH$_2$CHO, CH$_2$OHCHO,
HOCH$_2$CH$_2$OH, c--C$_2$H$_4$O, CH$_3$CHO, and HCCCHO in three typical GC molecular 
clouds. These molecules show
very high abundances relative to H$_2$, ranging from 10$^{-10}$ to
10$^{-8}$. 
The relative abundances of the different molecules with respect to that of CH$_3$OH are 
very similar within a factor of 3 for the three sources in the CMZ and likely 
everywhere in this region, as previously observed for a larger sample of sources
for other related COMs 
(Paper I). The new data strengthens our previous claim of a uniform grain mantle 
composition and similar chemical histories of the grains mantles in the CMZ clouds. 
However,
comparison between the CMZ clouds and hot cores shows very large differences in
the relative abundances of some of the molecules presented in this paper, 
indicating a different
grain mantle composition in the Galactic disk hot cores possibly due to a 
different chemical history. Hydrogenation and carbon addition seem to be more 
efficient in the GC 
than in the hot cores of the Galactic disk. The Galactic center
is the most extreme environment in the Galaxy with the presence of energetic 
processes, like UV radiation, X-rays and cosmic rays. These will affect the 
chemistry on the grain mantles in the CMZ and/or increase the H and C abundances
in the gas phase available for fast hydrogenation and carbon addition of unsaturated molecules on grain 
mantles. Other hypotheses, like the formation from radicals driven by cosmic
ray, seem also to be very promising. The CMZ is one of the most promising sources to increment our
knowledge of the chemical complexity in the interstellar medium.\\

\begin{figure}[h!]
\rotate
\centering
\includegraphics[angle=90,width=12cm,height=18cm]{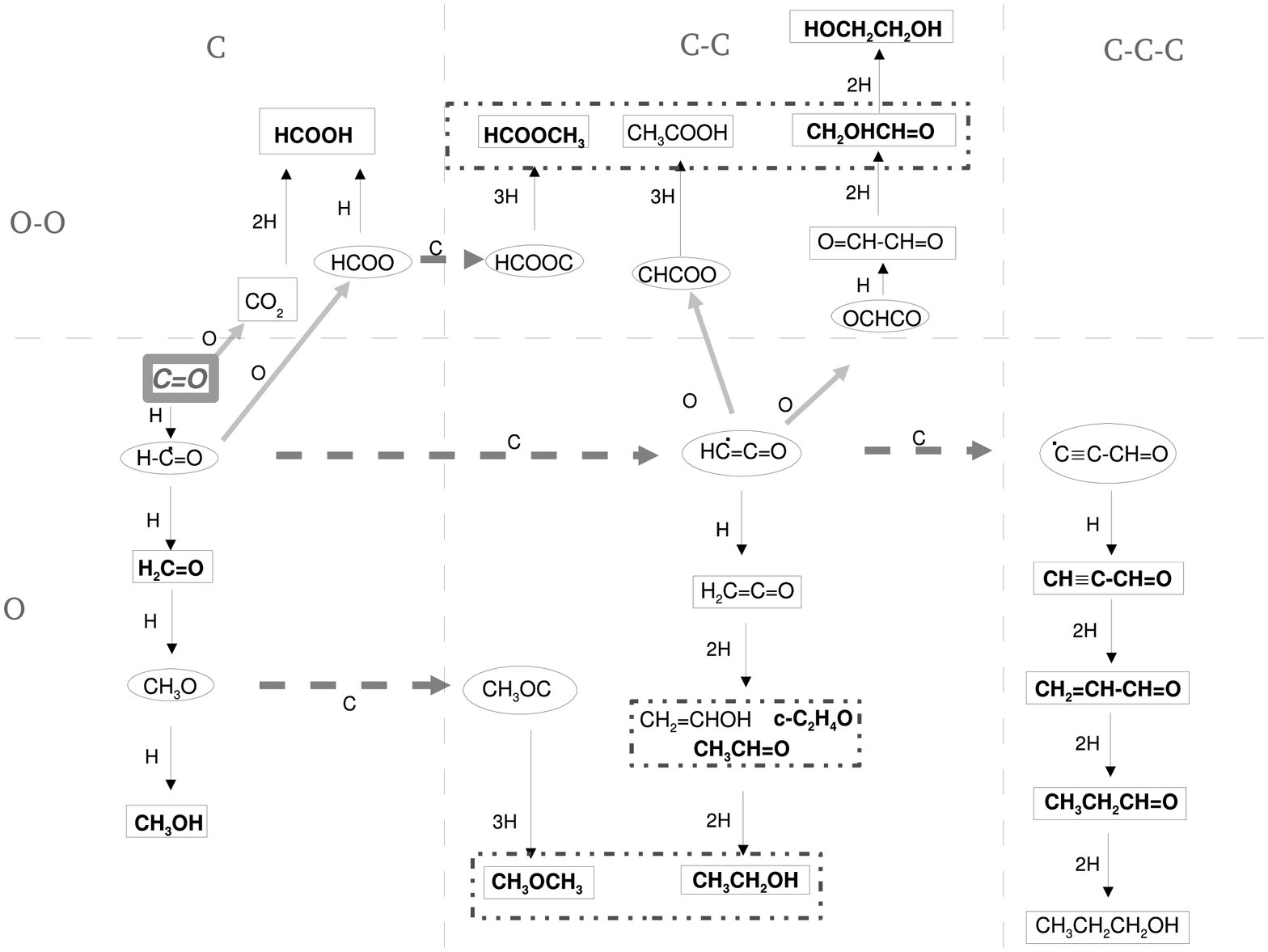}
\caption{
Proposed scheme for the formation of COMs on interstellar grain 
surfaces by addition of C, O and H on the
grain mantles. Complexity increases from left
to right by adding C (three areas C, C-C and C-C-C separated by dashed lines),
and from center to bottom and to top by adding H and
in diagonal by adding O (two areas O and O-O separated by dashed lines).
Radicals are encapsulated in elipses and the other molecular species in boxes.
 The molecules in boldface have been already detected in the GC
clouds studied by us. The other molecules have been not detected
or their spectroscopic parameters for their search are unkown. The thin dashed 
boxes separate the families of isomers. Adapted from 
\citet{cha05,tie82}.
\label{oxi}}
\end{figure}

\acknowledgments
We thank A. G. G. M. Tielens for his comments on Fig. \ref{oxi} and for a
discussion of the formation of COMs on grain mantles that contributed 
substantially to the improvement of  the final
version of the manuscript.
This work has been supported by the Spanish Ministerio de Educaci\'on y Ciencia 
under projects AYA~2003-02785, ESP~2004-00665, and
``Comunidad de Madrid" Government under PRICIT project S-0505/ESP-0237
(ASTROCAM). The participation of M. Morris in this project was supported by NSF
award INT-0340750 to UCLA.\\
{\it Facilities:} \facility{Green Bank Telescope}

\begin{deluxetable}{lccc}
\tabletypesize{\tiny}
\tablewidth{0pc}
\tablecolumns{7}
\tablecaption{observed frequencies.\label{freq}}
\tablehead{molecule&transition		&frequency		&E$_u$\\
	&	&(MHz)&(K)}
\startdata
 H$_2$COH$^+$		  &2$_{1,1}$$\rightarrow$2$_{1,2}$      &15532.99$^1$	  &17.49  \\ 
 CH$_3$CH$_2$CH$_2$OH * 	  &6$_{15}$$\rightarrow$6$_{0,6}$      &20339.80$^1$	  &10.30   \\
 CH$_2$OHCHO		  &1$_{1,0}$$\rightarrow$1$_{0,1}$  &13477.17$^1$   &1.20  \\
 CH$_2$OHCHO		  &2$_{1,1}$$\rightarrow$2$_{0,2}$  &15176.46$^1$   &2.38  \\
 CH$_2$OHCHO		  &4$_{1,3}$$\rightarrow$3$_{2,2}$  &15261.66$^1$   &6.48  \\
 CH$_2$OHCHO		  &4$_{1,3}$$\rightarrow$4$_{0,4}$  &22142.67$^1$   &6.48  \\
 HOCH$_2$CH$_2$OH &3$_{0,3}$(v0)$\rightarrow$2$_{0,2}$(v1)&23393.10$^1$ &2.92  \\
 HOCH$_2$CH$_2$OH &2$_{0,2}$(v0)$\rightarrow$1$_{0,1}$(v1)&13380.60$^1$ &1.47  \\
 HC$_2$CHO		  &2$_{0,2}$$\rightarrow$1$_{0,1}$      &18650.31$^3$   &1.34   \\
 HC$_2$CHO*		  &2$_{1,1}$$\rightarrow$1$_{1,0}$      &18978.78$^3$   &4.41   \\
 CH$_2$CHCHO		  &2$_{1,1}$$\rightarrow$1$_{1,0}$      &18221.16$^3$   &3.37   \\
 CH$_2$CHCHO		  &3$_{1,3}$$\rightarrow$2$_{1,2}$      &26079.45$^3$   &4.56  \\
 CH$_3$CH$_2$CHO	  &2$_{1,2}$$\rightarrow$1$_{1,1}$      &19690.43$^3$   &1.97   \\
 CH$_3$CH$_2$CHO	  &3$_{0,3}$$\rightarrow$2$_{1,2}$      &21451.57$^3$   &3.00   \\
 CH$_3$OH	  &3$_{2,3}$$\rightarrow$3$_{1,3}$	  &24928.70$^1$   &36.18  \\
 CH$_3$OH	  &4$_{2,3}$$\rightarrow$4$_{1,3}$	  &24933.47$^1$   &45.46   \\
 CH$_3$OH	  &2$_{2,3}$$\rightarrow$2$_{1,3}$	  &24934.38$^1$   &29.21   \\
 CH$_3$OH	  &5$_{2,3}$$\rightarrow$5$_{1,3}$	  &24959.08$^1$   &57.07  \\
 CH$_3$OH	  &6$_{2,3}$$\rightarrow$6$_{1,3}$	  &25018.12$^1$   &71.01  \\
 HCOOCH$_3$-A		  &2$_{1,1}$$\rightarrow$1$_{1,0}$	  &26048.52$^1$   &2.54   \\
 HCOOCH$_3$-E		  &2$_{1,1}$$\rightarrow$1$_{1,0}$	  &26044.83$^1$   &2.56   \\
 CH$_3$CHO-A *  	  &1$_{0,1}$$\rightarrow$0$_{0,0}$	  &19265.16$^2$   &0.93   \\
 CH$_3$CHO-E *  	  &1$_{0,1}$$\rightarrow$0$_{0,0}$	  &19262.16$^2$   &1.02   \\
 H$_2$CCO*		  &1$_{0,1}$$\rightarrow$0$_{0,0}$	  &20209.20$^1$   &0.97   \\
 c--C$_2$H$_4$O 	  &2$_{1,1}$$\rightarrow$2$_{0,2}$	  &24923.64$^1$   &6.15   
\enddata
\tablecomments{Transitions observed for each source. the spectroscopic
characteristics have been obtained from the Colonia University Database 
\citep[$^1$ ][]{mul01,mul05}, the Jet Propulsion Laboratory catalog \citep[$^2$
][]{pic98} and from \citet{hol04b} ($^3$). The transitions marked wit a * have
been only observed in MC~G+0.693--0.03.}
\end{deluxetable}

\begin{deluxetable}{lccccc}
\tabletypesize{\tiny}
\tablewidth{0pc}
\tablecolumns{7}
\tablecaption{Parameters of the sources.\label{poss}}
\tablehead{Source&$\alpha$[J2000]&$\delta$[J2000]&velocity&n(H$_2$)&X(CH$_3$OH)\\
&&&(km s$^{-1}$)&(cm$^{-2}$)&}
\startdata
MC~G--0.02--0.07 &17$^{\rm h}$45$^{\rm m}$51$^{\rm s}$&--28$^o$59$'$06$''$&47&6.8$\times$10$^{22}$&2.9$\times$10$^{-7}$\\
MC~G--0.11--0.08 &17$^{\rm h}$45$^{\rm m}$39$^{\rm s}$&--29$^o$04$'$05$''$&21&1.0$\times$10$^{22}$&1.1$\times$10$^{-6}$\\
MC~G+0.693--0.03 &17$^{\rm h}$47$^{\rm m}$22$^{\rm s}$&--28$^o$21$'$27$''$&68&4.1$\times$10$^{22}$&4.5$\times$10$^{-7}$ 
\enddata
\tablecomments{Positions, velocities, column  densities of H$_2$, and CH$_3$OH
abundances for the different sources. Data from Paper I.}
\end{deluxetable}

\begin{deluxetable}{lc|ccc|ccc|ccc}
\tabletypesize{\tiny}
\tablewidth{0pc}
\tablecolumns{7}
\tablecaption{Gaussian fit parameters.\label{gauss}}
\tablehead{Molecule&Transition& Intens. & V& $\Delta$V& Intens. & V& $\Delta$V&
Intens. & V& $\Delta$V\\
&  &   K&   km s$^{-1}$&   km s$^{-1}$
&   K&   km s$^{-1}$&   km s$^{-1}$
&   K&   km s$^{-1}$&   km s$^{-1}$}
\startdata
			&			&\multicolumn{3}{c}{\footnotesize MC~G--0.02}		&\multicolumn{3}{c}{\footnotesize MC~G$-$0.11}		& \multicolumn{3}{c}{\footnotesize MC~G+0.693} 	\\     
\hline
   H$_2$COH$^+$	&  2$_{1,1}$$\rightarrow$2$_{1,2}$	&   ...	&   ...	&   ...	&   ...	&   ...	&   ...	&  0.005(1)	&  73(3)  	&  22(6)\\
   CH$_3$CH$_2$CH$_2$OH	&  6$_{1,5}$$\rightarrow$6$_{0,6}$	&   ...	&   ...	&   ...	&   ...	&   ...	&   ...	&  $<$0.024	&   ...	&   ...\\
   CH$_2$OHCHO	&  1$_{1,0}$$\rightarrow$1$_{0,1}$	&  0.010(1)	&  52(1)	&  23(3) 	&  0.009(1)	&  22(1)  	&  21(3)  	&  0.023(1)	&  69(1)  	&  24(1)\\
   CH$_2$OHCHO	&  2$_{1,1}$$\rightarrow$2$_{0,2}$	&  0.004(1)	&  47(3)	&  17(5)	&  $<$0.006	&   ...	&   ...	&  0.013(1)	&  69(1)  	&  15(2)\\
   CH$_2$OHCHO	&  4$_{1,3}$$\rightarrow$3$_{2,2}$	&  0.005(1)	&  53(2)	&  19(4)	&  0.006(2)	&  22(2)	&  17(3)  	&  0.007(1)	&  67(2)  	&  21(5)\\
   CH$_2$OHCHO	&  4$_{1,3}$$\rightarrow$4$_{0,4}$	&  $<$0.015	&   ...	&   ...	&  0.008(1)	&  15(2)	&  21(6)	&  0.011(3)	&  71(3)	&  24(9)\\
   HOCH$_2$CH$_2$OH&  3$_{0,3}$(0)$\rightarrow$2$_{0,2}$(1)&  0.02(1)&  47(1)	&  20(2)	&  0.032(4)	&  18(1)  	&  17(1)  	&  0.060(6)	&  67(3)  	&  23(8)\\
   HOCH$_2$CH$_2$OH&  2$_{0,2}$(0)$\rightarrow$1$_{0,1}$(1)&  0.008(1)&  50(2)	&  15(4)	&  0.013(1)	&  18(1)  	&  14(2)  	&  0.020(1)&  59(1) 	&  24(2)\\
   HC$_2$CHO	&  2$_{0,2}$$\rightarrow$1$_{0,1}$   	&  0.007(2)	&  49(2)	&  19(3)	&  0.008(3)	&  15(2)	&  16(7)	&  0.019(1)	&  64(1)  	&  23(2)\\
   HC$_2$CHO	&  2$_{1,1}$$\rightarrow$1$_{1,0}$   	&   ...	&   ...	&   ...	&   ...	&   ...	&   ...	&  0.005(1)&  70(3)	&  21(10)\\
   CH$_2$CHCHO	&  2$_{1,1}$$\rightarrow$1$_{1,0}$   	&  0.004(1)	&  48(3)	&  20(5)	&  0.009(1)&  19(1)	&  14(3)	&  0.009(1)&  66(2)	&  23(3)\\
   CH$_2$CHCHO	&  3$_{1,3}$$\rightarrow$2$_{1,2}$   	&  $<$0.015 &   ...	&   ...	&  0.010(2)	&  20(1)	&  17(3)	&  0.013(5)	&  67(2)	&  15(3)\\
   CH$_3$CH$_2$CHO	&  2$_{1,2}$$\rightarrow$1$_{1,1}$   	&  0.006(1)	&  49(2)	&  15(5)	&  $<$0.006	&  ...	&   ...	&  0.006(2)	&  65(3)	&  21(6)\\
   CH$_3$CH$_2$CHO	&  3$_{0,3}$$\rightarrow$2$_{1,2}$   	&  $<$0.009	&   ...	&   ...	&  $<$0.009	&  ...	&   ...	&  0.006(3)	&  68(3)	&  29(6)\\
   CH$_3$OH	     	&  3$_{2,3}$$\rightarrow$3$_{1,3}$  	&  0.39(2)	&  46(1)	&  21(1)	&  0.24(2)	&  19(1)	&  16(3)	&  0.57(4)	&  69(1)	&  20(1)\\
   CH$_3$OH	     	&  4$_{2,3}$$\rightarrow$4$_{1,3}$	&  0.330(1)	&  46(1)	&  21(1)	&  0.177(1)	&  18(1)	&  16(1)	&  0.423(2)	&  69(1)	&  20(1)\\
   CH$_3$OH	     	&  2$_{2,3}$$\rightarrow$2$_{1,3}$	&  0.373(1)	&  46(1)	&  21(1)	&  0.226(1)	&  18(1)	&  16(1)	&  0.543(3)	&  69(1)	&  20(1)\\
   CH$_3$OH	     	&  5$_{2,3}$$\rightarrow$5$_{1,3}$	&  0.176(6)	&  45(1)	&  21(1)	&  0.097(6)	&  19(1)	&  16(7) 	&  0.19(2)	&  69(1)	&  21(1)\\
   CH$_3$OH	     	&  6$_{2,3}$$\rightarrow$6$_{1,3}$	&  0.118(6)	&  45(1)	&  22(1)	&  0.060(4)	&  19(1)	&  15(1)	&  0.101(8)	&  70(1)	&  24(1)\\
   HCOOCH$_3$-A	&  2$_{1,1}$$\rightarrow$1$_{1,0}$	&  0.020(6)	&  49(2)	&  27(6)	&  0.021(2)	&  19(1)	&  19(1) 	&  0.042(5)	&  65(1) 	&  25(2)\\
   HCOOCH$_3$-E	&  2$_{1,1}$$\rightarrow$1$_{1,0}$	&  0.018(6)	&  46(2)	&  23(5)	&  0.022(2)	&  17(1)	&  17(1)	&  0.041(5)	&  67(1)  	&  22(2)\\
   CH$_3$CHO-A 	&  1$_{0,1}$$\rightarrow$0$_{0,0}$	&   ...	&   ...	&   ...	&   ...	&   ...	&   ...	&  0.21(2)	&  67(1)	&  23(1)\\
   CH$_3$CHO-E 	&  1$_{0,1}$$\rightarrow$0$_{0,0}$	&   ...	&   ...	&   ...	&   ...	&   ...	&   ...	&  0.18(2)	&  67(1)	&  22(1)\\
   H$_2$CCO 	&  1$_{0,1}$$\rightarrow$0$_{0,0}$	&   ...	&   ...	&   ...	&   ...	&   ...	&   ...	&  0.043(6)	&  64(1)	&  17(3)\\
   c--C$_2$H$_4$O	&  2$_{1,1}$$\rightarrow$2$_{0,2}$	&  0.017(5)	&  51(1)	&  12(3)  	&  0.018(2) &  17(1)	&  17(2) 	&  0.032(4)	&  66(2)   	&  20(3)
\enddata
\tablecomments{The intensities are shown in antenna temperature units (K), and
the upper limits correspond to 3$\sigma$.}
\end{deluxetable}

\begin{deluxetable}{lcccccc}
\tabletypesize{\tiny}
\tablewidth{0pc}
\tablecolumns{7}
\tablecaption{Abundance ratios with respect to H$_2$.\label{abun}}
\tablehead{molecule&MC G--0.02		&MC G--0.11		&MC G+0.693}
\startdata
HC$_2$CHO	& 0.5E-9& 2.3E-9 &1.6E-9\\
CH$_2$CHCHO	& 0.3E-9& 2.3E-9 &0.9E-9\\
CH$_3$CH$_2$CHO	& 1.4E-9& 4.4E-9 &3.9E-9\\
CH$_2$OHCHO	& 0.3E-8& 1.8E-8 &0.9E-8\\
HOCH$_2$CH$_2$OH& 0.4E-8& 2.8E-8 &1.1E-8\\
HCOOCH$_3$	& 1.0E-8& 7.8E-8 &4.7E-8\\
c--C$_2$H$_4$O	& 1.1E-9& 5.6E-9 &3.0E-9\\
CH$_3$CHO$^*$	& 1.0E-8& 3.0E-8 &3.6E-8\\
H$_2$CCO$^*$	& 0.2E-8& 1.6E-8 &0.7E-8\\
H$_2$CO$^*$	& 1.6E-8& 6.5E-8 &0.9E-8\\
H$_2$COH$^+$$^*$& $<$1.1E-9& $<$1.3E-9& 2.4E-9\\
CH$_3$OH	&0.3E-6 &1.1E-6 &0.5E-6
\enddata
\tablecomments{Relative abundances with respect to that 
of H$_2$. 
The H$_2$ column densities used to estimate 
the abundances are
6.8$\times$10$^{22}$~cm$^{-2}$ for MC~G--0.02--0.07, 1$\times$10$^{22}$~cm$^{-2}$ for 
MC~G--0.11--0.08, and 4.1$\times$10$^{22}$~cm$^{-2}$ for MC~G+0.693--0.03, from
Paper I. 
For the molecules marked with a $^*$ we have used data from \citet{mar06}.}
\end{deluxetable}

\begin{deluxetable}{lcccccc}
\tabletypesize{\tiny}
\tablewidth{0pc}
\tablecolumns{7}
\tablecaption{Abundance ratios with respect to CH$_3$OH.\label{abun2}}
\tablehead{molecule&MC G--0.02		&MC G--0.11		&MC G+0.693}
\startdata
HC$_2$CHO	&1.6E-3		&2.1E-3		&3.6E-3	\\
CH$_2$CHCHO	&9.1E-4		&2.1E-3		&2.0E-3	\\
CH$_3$CH$_2$CHO	&3.9E-3		&$<$4.0E-3	&8.7E-3	\\
CH$_2$OHCHO	&1.0E-2		&1.5E-2		&2.0E-2	\\
HOCH$_2$CH$_2$OH&1.2E-2		&2.0E-2		&2.4E-2	\\
HCOOCH$_3$	&3.4E-2		&7.1E-2		&1.0E-1	\\
c--C$_2$H$_4$O	&3.7E-3		&5.1E-3		&6.7E-3	\\
CH$_3$CHO$^*$	&3.5E-2		&2.7E-2		&5.7E-2	\\
H$_2$CCO$^*$	&6.4E-3		&1.5E-2		&1.6E-2	\\
H$_2$CO$^*$	&5.3E-2		&5.9E-2		&2.0E-2	\\
H$_2$COH$^+$$^*$&$<$3.8E-3	&$<$1.2E-3	&5.3E-3	\\
C$_3$H$_7$OH	&\nodata	&\nodata	&$\le$1.6E-2	
\enddata
\tablecomments{Relative abundances of the different molecules with respect to that 
of CH$_3$OH. For the molecules marked with a $^*$ we have used data from
\citet{mar06}.} 
\end{deluxetable}

\begin{deluxetable}{lcccc}
\tabletypesize{\tiny}
\tablewidth{0pc}
\tablecolumns{5}
\tablecaption{Relative abundances between related molecules\label{fam}}
\tablehead{molecular ratios		&MC G--0.02		&MC G--0.11	
&MC G+0.693		&hot cores}
\startdata
\hline
Isomers\\
c--C$_2$H$_4$O:CH$_3$CHO		&0.1:1			&0.2:1			&0.1:1			&0.08--0.2:1\\
(CH$_3$)$_2$O$^1$:C$_2$H$_5$OH$^1$	&1.3:1			&0.9:1			&0.4:1			&3.2:1\\
CH$_3$COOH$^1$:CH$_2$OHCHO:HCOOCH$_3$	&$<$0.05:0.3:1		&$<$0.05:0.2:1		&$<$0.04:0.2:1		&$\sim$0.04:0.02:1\\
\hline
Aldehyde/alcohol\\
H$_2$CO:CH$_3$OH			&0.05:1			&0.06:1			&0.02:1			&0.2:1$^2$\\
CH$_3$CHO:C$_2$H$_5$OH			&0.9:1			&0.5:1			&0.8:1			&1.5E-3:1$^2$\\
CH$_3$CH$_2$CHO:C$_3$H$_7$OH		&\nodata		&\nodata			&$>$0.5:1		&\nodata\\
CH$_2$OHCHO:HOCH$_2$CH$_2$OH		&0.8:1			&0.8:1			&0.8:1			&0.5-1.4:1\\
\hline
Hydrogenation\\
H$_2$CCO:CH$_3$CHO			&1:5.6			&1:1.8			&1:3.6			&1:0.2$^2$\\ 
HC$_2$CHO:CH$_2$CHCHO:CH$_3$CH$_2$CHO	&1:0.6:2.4		&1:1.0:$<$1.9		&1:0.6:2.4		&1:0.8:4.7\\
\hline
H$_2$CO:H$_2$COH$^+$:CH$_2$OHCHO	&1:$<$0.07:0.2		&1:$<$0.02:0.3		&1:0.3:1		&1:$\sim$0.1:0.03
\enddata
\tablecomments{CMZ values for $^1$ from Paper I. 
The values of the Sgr~B2N hot core are obtained 
from: \citet{hol01,hol02,hol04b,num98,hal06,ohi96}. The values with $^2$ are from \citet{bis07} 
and are the averaged values of some hot cores, 
in their work the CH$_3$CHO and H$_2$CCO lie in a cooler gas than the C$_2$H$_5$OH, and the rest of the 
COMs.}
\end{deluxetable}


\end{document}